# Facile preparation of CuCo$_2$S$_4$/Cu$_7$S$_4$ nanocomposites as high-performance cathode materials for rechargeable magnesium batteries

Qin Zhang[1], Yaobo Hu[1,2,*], Jun Wang[1], Yuanxiao Dai[1], Fusheng Pan[1,2]

[1] College of Materials Science and Engineering, Chongqing University, Chongqing, China, 400044

[2] National Engineering Research Center for Magnesium Alloys, Chongqing, China, 400044

**ABSTRACT:** Searching for efficient and high-performance cathode materials for rechargeable magnesium ion batteries (RMBs) is urgent for exploring sustainable energy technologies. However, the majority of cathode materials for RMBs usually suffer from low rate capacity and inferior cycle performance caused by structure collapse. Herein, ternary CuCo$_2$S$_4$/Cu$_7$S$_4$ composites with nano-sized are first synthesized by a facile solvothermal method in this work and the magnesium ion storage behavior is discussed when applied in the cathode for RMBs. Electrochemical results demonstrate that the nanosphere-like CuCo$_2$S$_4$/Cu$_7$S$_4$ composites exhibit a high initial discharge capacity of 256 mAh g$^{-1}$ at 10 mA g$^{-1}$ and 123 mAh g$^{-1}$ at 300 mA g$^{-1}$ at room temperature. Furthermore, an outstanding long-term cyclic stability with a reversible capacity of 106 mAh g$^{-1}$ after 300 cycles and the coulombic efficiency of about 99% are achieved at 300 mA g$^{-1}$. Such remarkable electrochemical performance is owing to the nanosphere-like structure, ternary composition and pseudocapacitive storage behavior of $CuCo_2S_4/Cu_7S_4$. The results of this work indicate that the $CuCo_2S_4/Cu_7S_4$ nanocomposites synthesized by a facile and efficient method are promising cathodes for RMBs in energy storage/conversion application.

## 1. Introduction

With the widespread use of mobile portable devices and electric vehicles, there is an urgent need to develop an effective and sustainable energy conversion/storage system. Lithium-ion batteries (LIBs) are widely used due to their high energy density and well-rounded technology.[1-3] However, the high cost of LIBs and the problem of lithium dendrites have made researchers strive to explore higher reserves and safer alternatives—beyond lithium-ion batteries, such as magnesium-ion batteries (MIBs), sodium-ion batteries (SIBs), and zinc-ion batteries (ZIBs).[4-7] Especially, rechargeable magnesium batteries (RMBs) have attracted increasing attention for the large theoretical capacity (3833 mAh cm$^{-3}$ and 2205 mAh g$^{-1}$), low redox potential (-2.36 V vs. SHE), superior safety, abundant source and low cost of Mg anode.[7-9] Nevertheless, the commercial development of RMBs is still hampered for the lack of appropriate cathode materials.[10-11] So far, the reported cathode materials for RMBs mainly include Chevrel phase $M_xMo_6T_8$ (M=Metal, T=S, Se),[12-13] $MT_2$ (M=Mo, Ti, W, Cu, etc., T=S, Se),[14-19] $V_2O_5$,[20-22] $MnO_2$,[23-25] $MoO_3$,[26] $MgMSiO_4$ (M=Mn, Fe, Co, etc.),[27-29] $N_xM_2(PO_4)_3$ (M=Transition metal, N=Li, Na, etc.).[30-31] Among them, transition metal sulfides (TMSs) have been demonstrated as one of the most promising cathodes for RMBs due to its good electrical conductivity, less structural rigidity, and weak interaction with divalent cations during the cycling process. However, some problems such as poor cycle stability and sluggish diffusion dynamics still exist.[32-33]

In recent years, many transition metal binary sulfides such as CuS,[34-36] CoS,[37] and FeS[38] have been widely used as cathode materials for RMBs. Compared with binary transition metal sulfides, ternary TMSs have better electrochemical properties for their more redox active sites and higher electronic conductivity.[39] Recently, ternary transition metal sulfides such as $NiCo_2S_4$ and $CuCo_2S_4$ have been used in LIBs and SIBs.[40-42] And Blanc, et al[43] reported a low-temperature synthesis of $CuCo_2S_4$ which exhibited a specific capacity of 140 mAh $g^{-1}$ at a current density of 10 mA $g^{-1}$ at room temperature when used as a cathode material for RMBs. However, the combination of binary and ternary sulfides used as cathode material for RMBs has hardly been reported. In addition, nanotechnology is considered as one effective method to improve the diffusion kinetics of cathode.[44-46] The reduction of the particle size of the cathode material can not only shorten the diffusion path of the magnesium ion and obtain excellent diffusion kinetics, but also increase the contact area between the cathode material and the electrolyte and improve the utilization ratio of the electrode material, achieve a higher specific capacity.[47-48]

Therefore, we firstly synthesized the $CuCo_2S_4/Cu_7S_4$ nanospheres with a diameter about 150-200 nm by a facile solvothermal approach. When used as the cathode material for RMBs, the $CuCo_2S_4/Cu_7S_4$ nanocomposites delivery an ultra-high discharge specific capacity of 256 mAh $g^{-1}$ in the first cycle at 10 mA $g^{-1}$ and 123 mAh $g^{-1}$ at 300 mA $g^{-1}$ and an unprecedented cycling stability over 300 cycles at room temperature. Three main reasons for the excellent electrochemical performance of $CuCo_2S_4/Cu_7S_4$ nanocomposites used as the cathode material for RMBs are concluded as follows: (i) nanoscale $CuCo_2S_4/Cu_7S_4$ shortens the diffusion distance of magnesium ions, which is beneficial to obtain better diffusion kinetics; (ii) multi-component provides more active sites for electrochemical reactions, which improves the conductivity and electrochemical properties of the materials; (iii) compared with the traditional materials that dominated by slow diffusion control process, the electrode materials with a certain proportion of pseudocapacitive energy storage properties are more favorable to improve the rate performance of electrode materials[40]. In brief, a high performance cathode of $CuCo_2S_4/Cu_7S_4$ nanocomposites was synthesized for the first by a facile hydrothermal method, which may provide mew insight for the development of RMBs.

## 2. Experimental Section

**2.1 Synthesis of $CuCo_2S_4/Cu_7S_4$:** $CuCo_2S_4/Cu_7S_4$ nanocomposites were synthesized by a facile solvothermal method. Typically, $Cu(NO_3)_2\ 3H_2O$ (1.208 g) and $Co(NO_3)_2\ 6H_2O$ (2.9103 g) were dissolved in ethylene glycol (100 mL) under vigorous stirring for 5 min. Then, $Na_2S_2O_3$ (2.37 g) was added and stirring for another 30 min. Subsequently, the obtained mixture was sealed in a Telfon-lined autoclave (150 mL) and kept at 160°C for 10 h. After naturally cooling down to room temperature, the precipitate was collected by high-speed centrifugation and washed with deionized water and absolute ethanol for 6 times alternately. Finally, the black sample was obtained by drying at 60°C for 24 h.

**2.2 Materials characterizations:** The crystal structure of $CuCo_2S_4/Cu_7S_4$ nanocomposites was characterized by X-ray diffraction (XRD) with Cu Kα radiation (D/max 2500 PC). The morphologies and structures were observed by scanning electron microscopy (SEM, JEOL JSM-7800F) and transmission electron microscopy (TEM, Talos F200S). Energy dispersive spectrometer (EDS) elemental mappings were also conducted with Talos F200S microscopy. The specific surface and pore structures of $CuCo_2S_4/Cu_7S_4$ nanosphere were detected through Nitrogen adsorption-desorption measurements (Quadrasorb 2MP). Element and valence information of $CuCo_2S_4/Cu_7S_4$ nanocomposites were obtained through X-ray photoelectron spectroscopy (XPS, ESCALAB250Xi).

**2.3 Electrochemical measurements:** The cathode slurry was obtained by mixing and stirring the active materials ($CuCo_2S_4/Cu_7S_4$ nanocomposites), binder (polyvinylidene fluoride) and conductive agent (Super P) at a weight ratio of 7:2:1 for more than 12 h, which was then coated onto current collector of carbon paper and dried at 80 °C for 12 h under vacuum. The prepared electrode foil was punched into discs with a diameter of 12 mm. The mass loading of the active material within the whole material was about 2 mg cm$^{-2}$. Electrochemical performances of $CuCo_2S_4/Cu_7S_4$ nanocomposites were evaluated in CR2032 coin cell, in which the prepared electrodes as cathode, a glass microfiber film (Whatman, GF/A) as separator, a solution of 0.4 M $(MgPhCl)_2$-$AlCl_3$ in THF as the electrolyte and magnesium metal as anode were used. All cells were assembled in an argon-filled glove box ($H_2O$, $O_2$ < 1 ppm). Galvanostatic charge-discharge measurements were conducted in Neware battery test systems at various current density between 0.01 and 2.0 V (vs Mg/Mg$^{2+}$). Cyclic voltammetry (CV) measurements were performed on the CHI 660E electrochemical work station (ChenHua Instruments Co, China) at different scan rate (0.1/0.5/1.0/1.5/2.0/2.5/3.0/3.5mV s$^{-1}$). In this work, all the current density and specific capacities were calculated based on the active material in the $CuCo_2S_4/Cu_7S_4$ working electrode.

### 3. Results and discussion

### 3.1 Structure and morphology

The crystal structure and phase compositions of the prepared $CuCo_2S_4/Cu_7S_4$ nanocomposites were characterized by XRD (Figure 1a). All the diffraction peaks are consistent with Carrollite $CuCo_2S_4$ (JCPDS No. 42-1450) and $Cu_7S_4$ (JCPDS No. 22-0250), indicating that the prepared samples were composed of $CuCo_2S_4$ and $Cu_7S_4$ with no characteristic diffraction peaks of other impurities appear. The strong and sharp diffraction peaks demonstrate the high crystallinity of $CuCo_2S_4/Cu_7S_4$. In addition, the pore structure information of $CuCo_2S_4/Cu_7S_4$ nanosphere was tested by nitrogen adsorption-desorption measurement. As shown in Figure 1b, the BET surface area of $CuCo_2S_4/Cu_7S_4$ nanosphere is 7.714 $m^2\ g^{-1}$, and the BJH pore size distribution is noticed at around 3 nm. The extremely fine pores may be assigned to the voids between the adjacent nanoparticles, providing additional channels for the migration of $Mg^{2+}$ during electrochemical reactions.

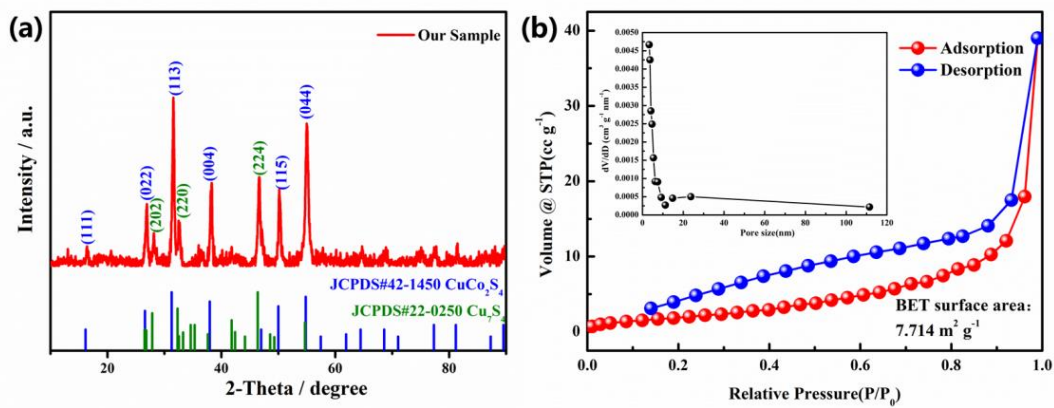

**Figure 1.** a) XRD pattern and b) $N_2$ adsorption–desorption pattern of $CuCo_2S_4/Cu_7S_4$ nanocomposites.

X-ray photoelectron spectroscopy (XPS) measurement was further performed to analyze the element composition and valence state distribution of the as-prepared $CuCo_2S_4/Cu_7S_4$ nanocomposites (Figure **2**). The survey spectrum of Figure 2a confirms the existence of Cu, Co and S elements. The O 1s signal peak is assigned to the exposure of the sample to air and the surface penetration depth of the XPS measurement. The high-resolution Cu 2p, Co 2p and S2p spectra of $CuCo_2S_4/Cu_7S_4$ are shown in Figure. 2b-d. Typically, the Cu 2p spectra (Figure 2b) contains two peaks at 932.3 and 952.3 eV, corresponding to Cu $2p_{3/2}$ and Cu $2p_{1/2}$, respectively. Moreover, two satellite peaks of $Cu^{2+}$ (denoted as "Sat.") are also observed at 932.8 and 953.1 eV. For the Co 2p spectrum (Figure 2c), two peaks at 779.0 and 794.0 eV are attributed to Co $2p_{3/2}$ and Co $2p_{1/2}$ of $Co^{3+}$, respectively. The two peaks at 781.4 and 797.4 eV are assigned to Co $2p_{3/2}$ and Co $2p_{1/2}$ of $Co^{2+}$, respectively.[49] The S 2p spectra shown in Figure 2d exhibits a satellite peak centered at 169.0 eV and another three peaks. The peaks at 161.4 and 162.5 eV are attributed to S $2p_{3/2}$ and $2p_{1/2}$, respectively, and the peak at 163.6 eV is corresponding to the bonding between the metals and sulfur (denoted as S-M bond).[50-51] In conclusion, the XPS results provide further evidence for the composition of $CuCo_2S_4/Cu_7S_4$ nanocomposites.

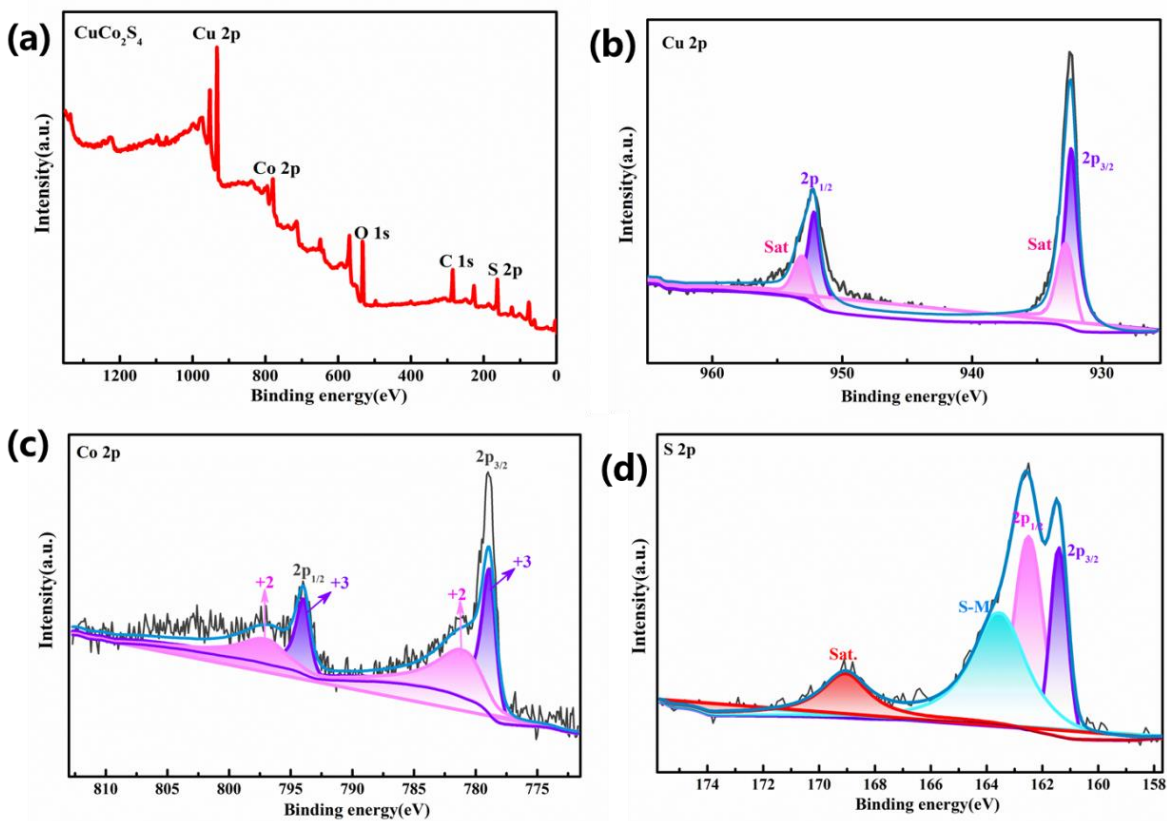

**Figure 2.** a) XPS survey spectra, b) Cu 2p, c) Co 2p and d) S 2p of $CuCo_2S_4/Cu_7S_4$ nanocomposites.

Morphology and structure characterization of $CuCo_2S_4/Cu_7S_4$ nanocomposites were examined by SEM and TEM as shown in Figure 3. The prepared $CuCo_2S_4/Cu_7S_4$ sample displays a number of uniformly dispersed nanoparticles (Figure 3a-b). It can be further demonstrated from the TEM image (Figure 3c) that the nanocomposite are composed of nanoparticles with a diameter of about 150-200 nm. HRTEM image (Figure 3d) shows two distinct lattice fringes with spacing of 0.288 nm and 0.322 nm, which can be assigned to the (113) plane of Carrollite $CuCo_2S_4$ and the (202) plane of $Cu_7S_4$, respectively, indicating the formation of $CuCo_2S_4/Cu_7S_4$ nanocomposites. The selected-area electron diffraction (SAED) pattern of Figure 3e suggests the characteristics of diffraction rings, indicating that the typical polycrystalline structure of $CuCo_2S_4/Cu_7S_4$ nanocomposites. The observed diffraction rings can be assigned to the (113) and (044) planes of the $CuCo_2S_4$, and (202) and (224) planes of the $Cu_7S_4$, which provides further powerful evidence for the synthesis of $CuCo_2S_4/Cu_7S_4$ nanocomposites. These results were in good agreement with the XRD results shown in Figure 1. In addition, the STEM image and the corresponding EDS elemental mapping analysis presented in Figure 3f-i visually reveals the homogenous dispersion of Cu, Co and S components in $CuCo_2S_4/Cu_7S_4$ nanocomposites.

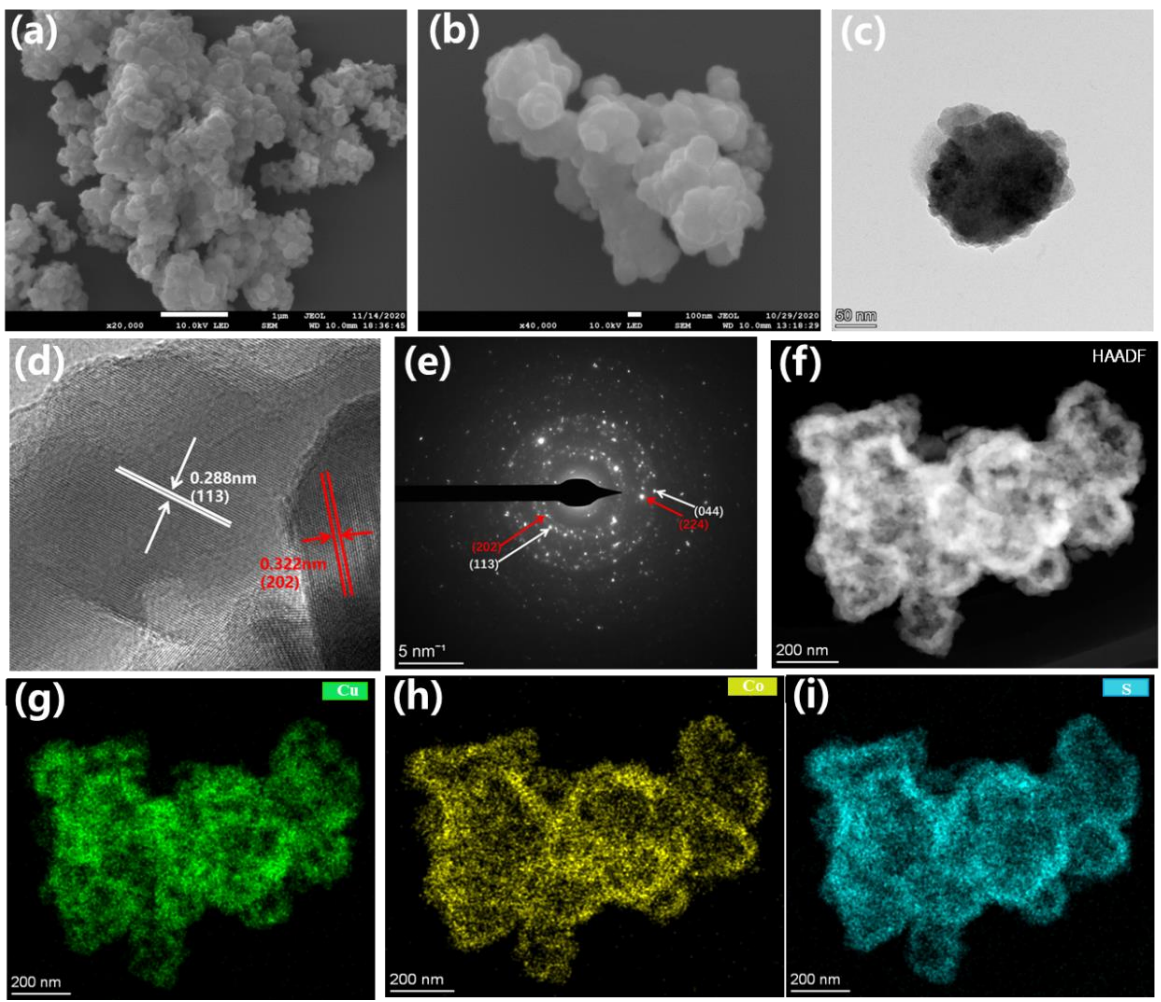

**Figure 3.** Morphology and structure characterization of $CuCo_2S_4/Cu_7S_4$ nanosphere: a-b) SEM images, c) TEM image, d) HRTEM image and e) the corresponding SAED pattern, f) STEM image and g-i) corresponding Cu, Co and S element mapping images.

## 3.2 Electrochemical performance

Figure 4 shows the electrochemical performance of $CuCo_2S_4/Cu_7S_4$ nanocomposites as cathode material for RMBs. In order to investigate the charge/discharge behaviors of $CuCo_2S_4/Cu_7S_4$ nanocomposites, cyclic voltammetry (CV) curves were measured at a scan rate of 0.1 mV s$^{-1}$ from 0.01-2.0V (Figure 4a). In the first cycle of $CuCo_2S_4/Cu_7S_4$ nanocomposites, the strong cathodic peaks located at around 1.25 V and 0.92 V may be related to the intercalation of $Mg^{2+}$ into the $CuCo_2S_4/Cu_7S_4$ lattices and the reduction of $Cu^{2+}$ and $Co^{3+}$ to Cu and Co. The anodic peaks centered at 1.53 V and 1.75 V are resulted from the possible formation of metal sulfides ($CuS_x$ or $CoS_x$).[40, 42] In the subsequent cycles, the anodic peak and the cathodic peak at 0.92 V shift slightly towards higher potential, and the cathodic peak at 1.25 V moves to lower potentials. The weak peak at 0.78 V that exists only in the first cycle could be attributed to the formation of solid electrolyte inter phase (SEI). All of the cathodic and anodic peaks at the CV curves after the first cycle are almost overlapped, demonstrating the outstanding reversibility for the $Mg^{2+}$ insertion/extraction process in $CuCo_2S_4/Cu_7S_4$ nanocomposites.

According to the galvanostatic charge-discharge curves in Figure 4b-e at different current density of 50, 200, 300 and 500 mA $g^{-1}$, it can be found that all typically charge-discharge curve platforms correspond well to the redox peaks in the CV curve mentioned above. Furthermore, voltage platform of the charge-discharge curve become shorter (lower capacity) after the first cycle, but also higher in polarization potential, which means that slow diffusion kinetics and large internal resistance within the cell. However, with the cycle progresses, the voltage platform essentially returns to its original length and a decrease in polarization (black arrows) occurs with the charge voltage decrease and discharge voltage increase, which assigns to an activation process. More interestingly, the activation degree of the $CuCo_2S_4/Cu_7S_4$ electrode at high current density is much higher than that at lower current density. After 240 cycles of activation, the discharge specific capacity at 500 mA $g^{-1}$ reaches 55 mAh $g^{-1}$, which is much higher than the first discharge specific capacity of 35 mAh $g^{-1}$. The possible reasons are concluded as: for lower current density, a greater depth of material is involved per cycle and therefore requires fewer cycles to recover. Conversely, as the current density increases and the surface reaction is the main control step, more active sites need to be activated with more cycles. The activation process is also observed by the cycling test. As shown in Figure 4f, the reversible capacity of $CuCo_2S_4/Cu_7S_4$ cathode at 200 mA g$^{-1}$ gradually increases and finally reaches a maximum of 111 mAh g$^{-1}$ at 160 cycles.

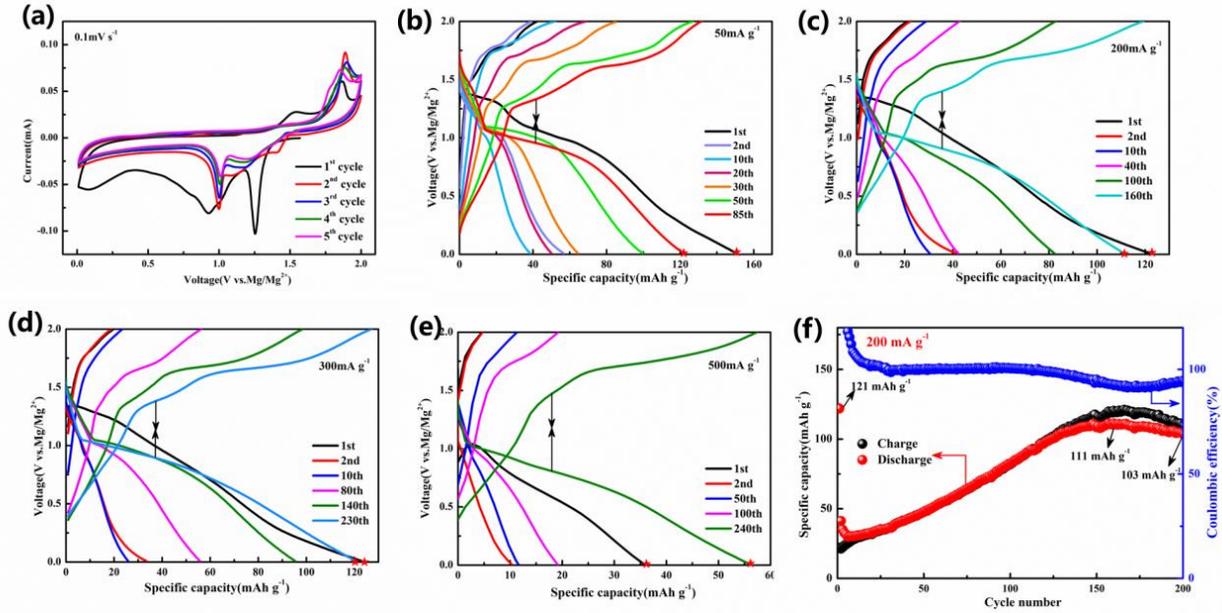

**Figure 4.** a) CV curves of the $CuCo_2S_4/Cu_7S_4$ electrode at a scan rate of 0.1mV s$^{-1}$. b-e) Galvanostatic charge-discharge profiles for $CuCo_2S_4/Cu_7S_4$ at different cycles at 50, 200, 300 and 500 mA g$^{-1}$. f) Cycling performance and Coulombic efficiency at 200 mA g$^{-1}$.

The rate capability of $CuCo_2S_4/Cu_7S_4$ nanocomposites was tested at different current densities (Figure 5a, b). As exhibited in Figure 5a, the $CuCo_2S_4/Cu_7S_4$ cathode exhibits initial discharge specific capacity about 256, 173, 150, 137, 121 and 123 mAh g$^{-1}$ at 10, 20, 50, 100, 200 and 300 mA g$^{-1}$, respectively. Moreover, the specific capacity of 35 mAh g$^{-1}$ can still obtained when the current density was increased up to 500 mA g$^{-1}$. Benefiting from the ternary composition and the nanosphere structure, $CuCo_2S_4/Cu_7S_4$ delivers reversible discharge capacities of about 105, 104, 96, 80, 74 and 67 mAh g$^{-1}$ at 10, 20, 50, 100, 200 and 300 mA g$^{-1}$, respectively (Figure 5b). When the current density was adjusted back to 10 mA g$^{-1}$, a higher reversible specific capacity of 111 mAh g$^{-1}$ can be obtained, indicating a superior tolerance of the host material to the current mutation. This can rarely be found in TMSs cathode materials for RMBs, demonstrating the potential application prospects for commercial RMBs.

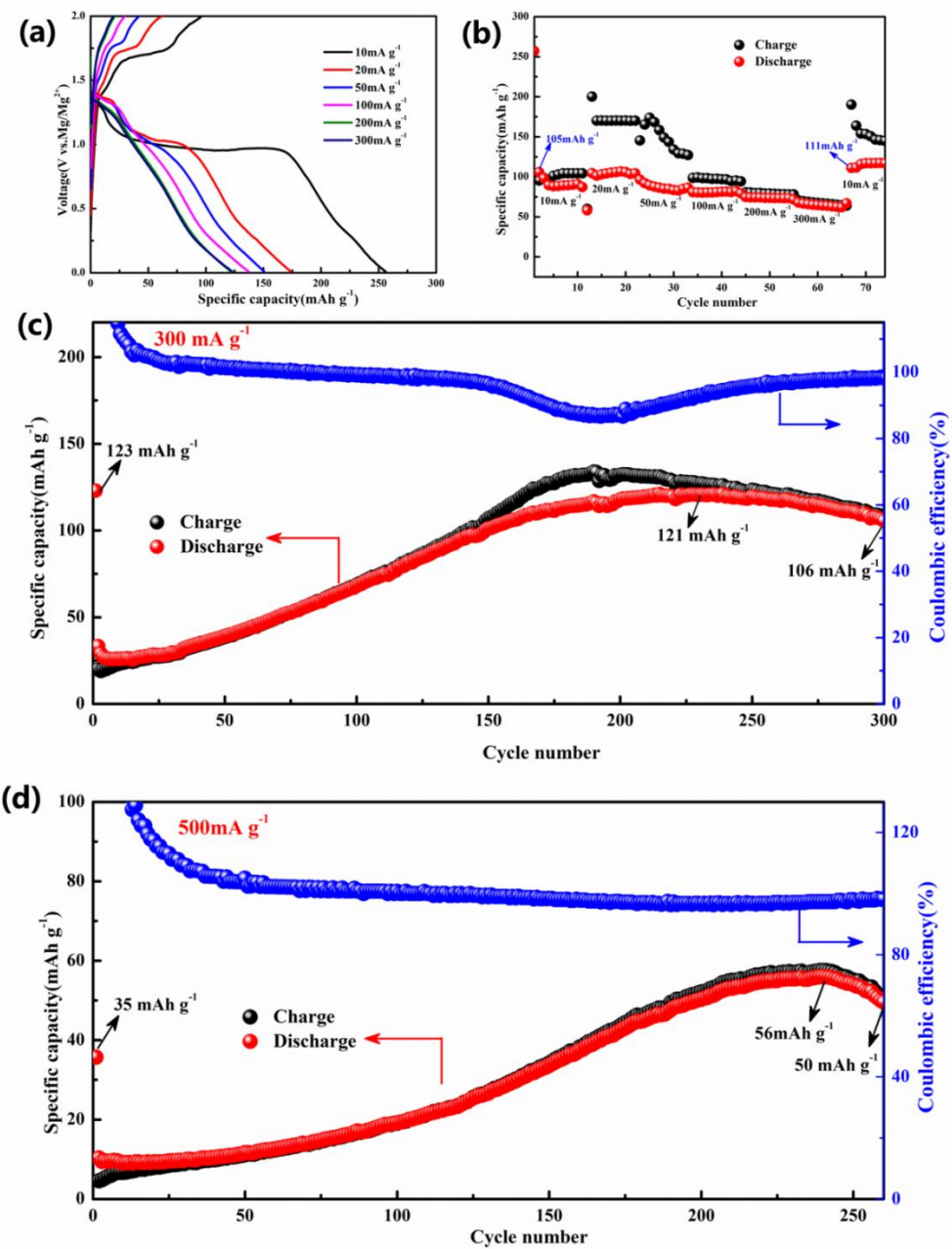

**Figure 5.** a) Galvanostatic charge-discharge profiles of $CuCo_2S_4/Cu_7S_4$ cathode at different current densities. b) Rate capacities of the $CuCo_2S_4/Cu_7S_4$ electrode with various current densities from 10 to 200 mA $g^{-1}$. c-d) Cycling performance of the $CuCo_2S_4/Cu_7S_4$ electrode at 300 and 500 mA $g^{-1}$.

The long-term cycling stability of $CuCo_2S_4/Cu_7S_4$ nanocomposite was performed at 300 and 500 mA $g^{-1}$. As is shown in Figure 5c, when used as cathode material of RMBs, the $CuCo_2S_4/Cu_7S_4$ nanocomposite exhibits outstanding cycling stability with a discharge specific capacity of 106 mAh $g^{-1}$ after 300 cycles (capacity retention rate of 86%) at 300 mA $g^{-1}$. What's more, the capacity remains 55 mAh $g^{-1}$ after 250 cycles even at 500 mA $g^{-1}$, much better than the initial discharge capacity (35 mAh $g^{-1}$). All the cycle curves show the same trend: the capacity decrease rapidly after the first cycle due to irreversible reactions and then gradually increases until it reaches stability. It can be observed that after a certain number of cycles there is a slightly decrease in the Coulomb efficiency of the electrode material at 200 (Figure 4f) and 300 mA $g^{-1}$ (Figure 5c), and then it returns to nearly 100%. However, the coulomb efficiency is almost certainly 100% at a higher current density of 500 mA $g^{-1}$ (Figure 5d). It can be assumed that the electrode experiences internal structure modification during the activation process, however, this change can be recovered after a certain cycle at low current density. While the surface control is the main reaction mode at the high current density of 500 mA $g^{-1}$ with faster kinetics, leaving little opportunity for internal structure changes. In conclusion, the faster kinetics provided by the nanometer size, the more reaction sites for the $CuCo_2S_4/Cu_7S_4$ nanocomposite by the multicomponent, and the possible extra capacity for $CuCo_2S_4$ by the formation of $Cu_7S_4$ all contribute to the high rate capacity, outstanding cycling properties, and excellent rate performance.

The excellent rate performance of CuCo$_2$S$_4$/Cu$_7$S$_4$ nanocomposites might not be explained by the slow magnesiation/demagnesiation process dominated by diffusion control. Considering the nanoscale and ternary composition characteristics of the CuCo$_2$S$_4$/Cu$_7$S$_4$, the outstanding rate performance could be explained by the pseudocapacitive effect based on surface control instead of the slow magnesiation/demagnesiation process dominated by diffusion control. To further examine the proportion of diffusion control and pseudocapacitive contribution, CV tests at different scan rates from 0.5 to 3.5 mV s$^{-1}$ were performed (Figure 6a). All the redox peaks correspond to the reaction processes of magnesium ions in the electrode material, including the Faraday reaction and the pseudocapacitive effect. Typically, the different electrochemical reaction mechanisms can be identified by the following equation,[52-53]

$$i = av^b \quad (1)$$

where the $i$ is the current value in the CV curves, $v$ is the scan rate, $a$ and $b$ are adjustable values determined by the log($i$)-log($v$) plot (Figure 6b). If the value of $b$ is about 0.5, the electrochemical reaction is dominated by diffusion-control. Whereas, the electrochemical reaction process is mainly attribute to surface-control when the value of $b$ is close to 1.0.[54-55] As shown in Figure 6b, the $b$-values obtained by linearly fitting the scatter plots are 0.795, 0.970 and 0.748 from peak 1 to peak 3, respectively, demonstrating that the storage process of magnesium ion in the CuCo$_2$S$_4$/Cu$_7$S$_4$ is mainly driven by strong pseudocapacitive effects.

The pseudocapacitive-contributions at various scan rates can be calculated by the Trasatti's expression[52-53]

$$i = k_1 v + k_2 v^{1/2} \quad (2)$$

where $k_1 v$ and $k_2 v^{1/2}$ represent the capacitive and diffusion-controlled contributions, respectively. The equation 2 can be further converted into equation 3.

$$\frac{I_p}{v^{0.5}} = k_1 v^{0.5} + k_2 \qquad (3)$$

the values of $k_1$ and $k_2$ were calculated by making plots of $I_p/v^{0.5}$-$v^{0.5}$ (Figure 6c), and then the capacitive contributions at different scan rates shown in Figure 6d can be obtained. It can be seen that the contribution of pseudocapacitance increases from 76.2% at 0.5mV s$^{-1}$ to a maximum of 93.7% at 2.0 mV s$^{-1}$ as the scan rates increases and finally achieves stability. This non-linear trend is mainly due to the fact that pseudocapacitive effect is not an inherent property of the material. The main reasons for the pseudocapacitive storage of $CuCo_2S_4/Cu_7S_4$ nanocomposites are the nanoscale and high conductivity of the ternary material. The above analysis determines that the capacity of the sample is mainly derived from the pseudocapacitive storage, and therefore contributed to the high specific capacity, excellent rate properties and outstanding cycling performance.

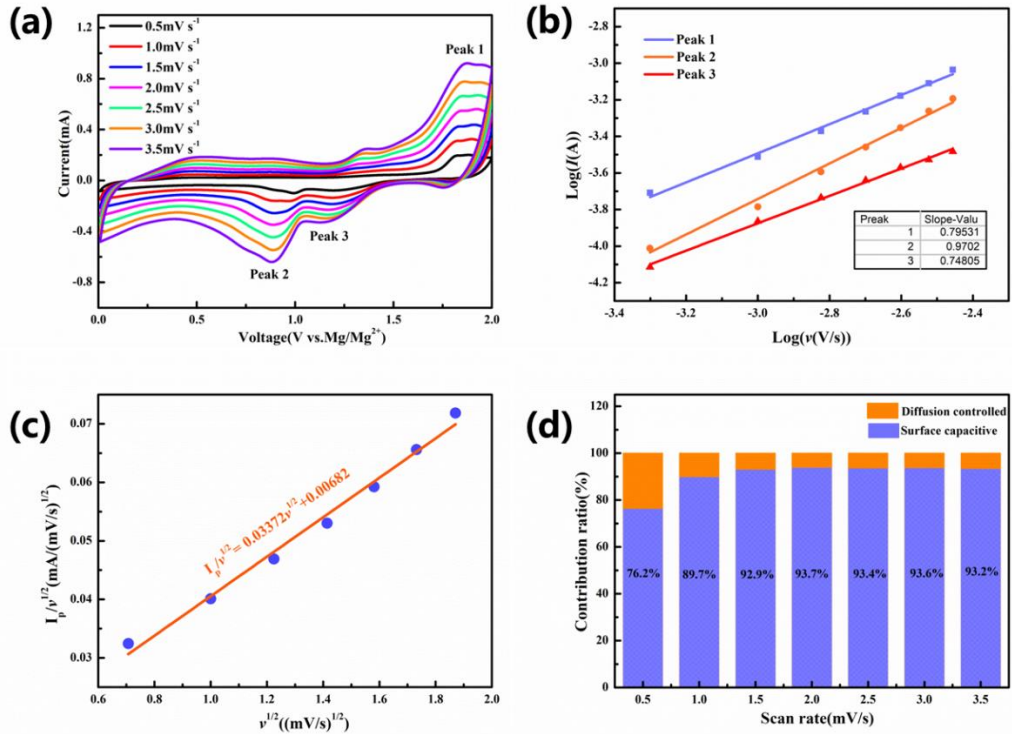

**Figure 6.** a) CV curves of CuCo$_2$S$_4$/Cu$_7$S$_4$ nanosphere at different scan rate; b) log($i$)-log($v$) relationship at different redox peaks; c) $I_p/v^{1/2}$-$v^{1/2}$ plot of CuCo$_2$S$_4$/Cu$_7$S$_4$; d) contribution ratio of pseudocapacitive at different scan rate.

To further reflect the potential applications visually, the battery is assembled with magnesium foil for the anode and CuCo$_2$S$_4$/Cu$_7$S$_4$ nanocomposites for the cathode. The fresh battery after aging for 24 hours can power and light the red LED (Figure **7**).

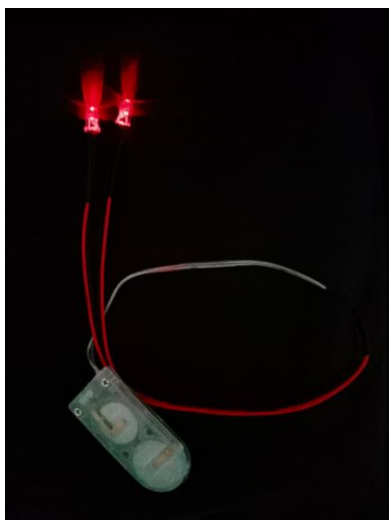

**Figure 7.** The as-assembled coin cells can light red LED.

### 4. Conclusions

In summary, ternary CuCo$_2$S$_4$/Cu$_7$S$_4$ nanocomposites were firstly fabricated by a facile solvothermal method and confirmed as an efficient cathode material for RMBs. Benefiting the nanosphere-like structure, the ternary composition and the pseudocapacitive effect, the CuCo$_2$S$_4$/Cu$_7$S$_4$ nanocomposites can deliver a high initial discharge capacity of 256 mAh g$^{-1}$ at 10 mA g$^{-1}$, a remarkable rate capacity of 123 mAh g$^{-1}$ at 300 mA g$^{-1}$ and an outstanding long-term cycling performance of 106 mAh g$^{-1}$ at 300 mA g$^{-1}$ after 300 cycles. Furthermore, the electrochemical storage mechanism demonstrated that the storage process of magnesium ion in the CuCo$_2$S$_4$/Cu$_7$S$_4$ cathode is mainly driven by the strong pseudocapacitive effects which promote the electrochemical reaction kinetics. All results confirmed that the ternary CuCo$_2$S$_4$/Cu$_7$S$_4$ nanocomposites have tremendous commercial potential in cathode for RMBs.

## AUTHOR INFORMATION

**Corresponding Author**

Yaobo Hu—National Engineering Research Center for Magnesium Alloys, Chongqing University, Chongqing 400044, China; E-mail: yaobohu@cqu.edu.cn

**Author Contributions**

The manuscript was written through contributions of all authors. All authors have given approval to the final version of the manuscript.

**Notes**

The authors declare no competing financial interest.